
\magnification\magstep1
\hsize 15.5truecm
\scrollmode
\overfullrule=0pt

 at 17.28truept
 at 14.4truept
 at 12truept
 at 10.95truept
 at 17.28truept
 at 10truept

\def\title#1{%
\vskip0pt plus.3\vsize\penalty-100%
\vskip0pt plus-.3\vsize\bigskip\vskip\parskip%
\bigbreak\bigbreak\centerline{\bf #1}\bigskip%
}

\def\chapter#1#2{\vfill\eject
\centerline{\bf Chapter #1}
\vskip 6truept%
\centerline{\bf #2}%
\vskip 2 true cm}

\def\section#1#2{%
\def\\{#2}%
\vskip0pt plus.3\vsize\penalty-100%
\vskip0pt plus-.3\vsize\bigskip\vskip\parskip%
\par\noindent{\bf #1\hskip 6truept%
\ifx\empty\\{\relax}\else{\bf #2\smallskip}\fi}}

\def\subsection#1#2{%
\def\\{#2}%
\vskip0pt plus.3\vsize\penalty-20%
\vskip0pt plus-.3\vsize\medskip\vskip\parskip%
\def\TEST{#1}%
\noindent{\ifx\TEST\empty\relax\else\bf #1\hskip 6truept\fi%
\ifx\empty\\{\relax}\else{#2\smallskip}\fi}}

\def\proclaim#1{\medbreak\begingroup\noindent{\bf #1.---}\enspace\sl}

\def\endproclaim{\endgroup\par\medbreak}

\def\qqbox#1{\quad\hbox{#1}\quad}


\def\comfig#1#2\par{
\medskip
\centerline{\hbox{\hsize=10cm\eightpoint\baselineskip=10pt
\vbox{\noindent #1}}}\par\centerline{ Figure #2}}

\def\figcom#1#2\par{
\medskip
\centerline
{Figure #1}
\par\centerline{\hbox{\hsize=10cm\eightpoint\baselineskip=10pt
\vbox{\noindent #2}}}}
\def\bull{~\vrule height .9ex width .8ex depth -.1ex}

\def\adresse#1{%
\bigskip\hfill\hbox{\vbox{%
\hbox{#1}\hbox{Centre de Math\'ematiques}\hbox{Ecole Polytechnique}%
\hbox{F-91128 Palaiseau Cedex (France)}\hbox{\strut}%
\hbox{``U.A. au C.N.R.S. n$^{\circ}$169''}}}}


\def\comfig#1#2\par{
\medskip
\centerline{\hbox{\hsize=10cm\eightpoint\baselineskip=10pt
\vbox{\noindent{\sl  #1}}}}\par\centerline{{\bf Figure #2}}}

\def\figcom#1#2\par{
\medskip
\centerline
{{\bf Figure #1}}
\par\centerline{\hbox{\hsize=10cm\eightpoint\baselineskip=10pt
\vbox{\noindent{\sl  #2}}}}}

\def\em{\sl}

\def\\{\hfill\break}

\long\def\adresse#1{%
\leftskip=0truecm%
\vskip 3truecm%
\hbox{\hskip 10.5truecm{\hsize=7.5truecm\vbox{%
\def\cr{\parskip0pt\par\noindent}\noindent#1}}}}

\def\bull{~\vrule height .9ex width .8ex depth -.1ex}

\def\a{\alpha}
\def\b{\beta}
\def\g{\gamma}

\def\d{\delta}
\def\D{\Delta}
\def\e{\varepsilon}

\def\la{\lambda}
\def\La{\Lambda}
\def\s{\sigma}

\def\f{\varphi}

\def\o{\omega}

\def\CC{{\bf C}}

\def\RR{{\bf R}}

\def\ZZ{{\bf Z}}\def\cA{{\cal A}}
\def\cB{{\cal B}}\def\cC{{\cal C}}\def\cD{{\cal D}}

\def\tr{\mathop{\rm tr}\limits}


\def\la{\lambda}

\def\sqr#1#2{{\vcenter{\hrule height.#2pt%
\hbox{\vrule width.#2pt height#1pt\kern#1pt%
\vrule width.#2pt}%
\hrule height.#2pt}}}

\def\ln{{\rm ln}}\def\Arg{{\rm Arg}}\def\Re{{\rm Re}}\def\Imm{{\rm
Im}}\def\Ad{{\rm Ad}}

\centerline{\bf Soliton solutions of the classical lattice sine-Gordon
system}
\bigskip
\centerline{B. Enriquez}
\medskip
{\bf Abstract. }{\em
We study the soliton-type solutions of the system introduced by B.
Feigin and the author in [EF]. We show that it reduces to a top-like
system, and we study the behaviour of the solutions at the lattice
infinity. We compute the scattering of the solitons and study some
periodic solutions of the system.}
\medskip
\section{}{Introduction.} In this work, we study solutions to the
lattice sine-Gordon system introduced by B.L. Feigin and the author in
[EF]. This system consists of two families of compatible flows,
analogues of the sine-Gordon and mKdV flows. We recall that the
solutions studied in the continuous context
are characterized by the requirement that they be stationary with
respect to a linear combination of those flows. We study the analogous
problem here, and reduce it to a top-like system. We then show
how to pass from the group variables satisfying this system to the
local variables (inverse scattering)~; in particular, we find
solutions with variables tending to a constant at lattice infinity
(soliton solutions). We then study the scattering of these solutions,
and we find a purely elastic behaviour. Two features are rather different
from the continuous case~: the absence of charge of the solitons and their
non-fermionic character (solutions with multiple poles exist). This is due to
the difference between the relevant subalgebras of
$\widehat{s\ell}_{2}$: positive Cartan and negative principal subalgebras
instead of positive and negative principal subalgebras
in the continuous case.

Finally, we study some periodic solutions of the system. We find that such
solutions are characterized by the condition that the difference of the
two points at infinity on the spectral curve is torsion in its Jacobian.

We express our thanks to B. Feigin for discussions related to this
problem and to Mme Harmide for typing this text.

\section{1.}{Review of continuous sine-Gordon system.} Let $\phi$ be a
function on $\RR$ and let us consider the matrix $M_{x} (\la)= P \exp
\int^{x}_{-\infty} (\La + \phi h). \bar n_{x}$, with $\bar n_{x}= n_{x}
\exp \mathop{\sum}\limits_{i>0 \atop {\rm odd}} \int_{x}^{\pm \infty}
u_{i}\La^{-i}$, and
$\partial_{x} + \La + \phi h = n_{x} (\partial + \La +
\mathop{\sum}\limits_{i>0\atop
{\rm odd}} u_{i} \La^{-i})  n^{-1}_{x}$, $h= \pmatrix{1 & 0\cr 0& -1\cr} , \La
=
\pmatrix{0 & 1\cr \la & 0\cr}$ ([DS])~; $P\exp \int^{x}_{-\infty}(\La
+\phi h)$ is developed around $\la=0$ and takes the form
$$
\pmatrix{1 + \la \int_{-\infty}^{x} e^{2\f(y)} \int_{-\infty}^{y}
e^{-2\f}+\cdots  & \int^{x}_{-\infty}e^{2\f}+\cdots\cr \la
\int^{x}_{-\infty} e^{-2\f}+ \cdots  &  1 + \la
\int^{x}_{-\infty}e^{-2\f(y)} \int_{-\infty}^{y} e^{2\f}+\cdots\cr}
\pmatrix{e^{\f(x)} & 0\cr 0 & e^{-\f(x)}} \ .
$$
Then the mKdV flows are $\partial^{KdV}_{n} M_{x} =
M_{x}\La^{2n+1},n\geq 0$ and the sine-Gordon flow is $\partial^{SG}_{0}
M_{x} = \La^{-1} M_{x}$. We introduce higher sine-Gordon flows by
$\partial_{n}^{SG} M_{x}= \La^{-(2n+1)} M_{x}$. We have
$$
\partial_{x}\partial_{0}^{SG} \f(x)= e^{2\f(x)} -e^{-2 \f(x)},
$$
$$
\partial_{x} \partial_{1}^{SG}\f(x) = e^{2\f(x)} \int^{x}_{-\infty}
e^{-2\f(y)} \int^{y}_{-\infty} (e^{2\f} -e^{-2\f}) -e^{-2\f(x)}
\int^{x}_{-\infty} e^{2\f(y)} \int^{y}_{-\infty} (e^{2\f} -e^{-2\f}),
$$
$$
\partial_{x} \partial_{n}^{SG} \f(x)= e^{2\f(x)} \int_{-\infty}^{x}
e^{-2\f(y)}  \partial_{y} \partial_{n-1}^{SG} \f (y)-e^{-2\f (x)}
\int^{x}_{-\infty} e^{2\f(y)} \partial_{y} \partial_{n-1}^{SG} \f(y).
$$
These equations are well defined for functions satisfying the boundary
conditions $e^{4\f(x)} \mathop{\longrightarrow}\limits_{x\to -\infty} 1$
faster than any inverse polynomial, and they preserve these boundary
conditions.
If $\f$ is such that in addition $e^{4\f(x)}
\mathop{\longrightarrow}\limits_{x\to +\infty}1$ faster than any inverse
polynomial, and if the conditions
$$
\int^{\infty}_{-\infty} (e^{2\f}
-e^{-2\f}) = \int^{\infty}_{-\infty} (e^{2\f(x)} \int^{x}_{-\infty}
e^{-2\f} -e^{-2\f(x)} \int^{x}_{-\infty} e^{2\f}) = \cdots =0
$$
are met (they are
a rewriting of $[\La , P\exp \int^{\infty}_{-\infty}(\La +\phi h)]=0$),
then the higher SG equations can be rewritten with $+\infty$
replacing $-\infty$, at point $\f$~; these conditions are preserved by
the flows.
[Note that any solution to this system, such that $e^{4\f}
\to 1$ and $\partial_{k}^{SG} \f \to 0$, faster than any ${1\over x^{N}}$
when $x\to \pm \infty$ will also satisfy the conditions
$\int^{\infty}_{-\infty} (e^{2\f} -e^{-2\f})=\cdots =0$, since we deduce from
$\partial_{k}^{SG} (\f h) = (\Ad^{-1}(P\exp
\int^{x}_{-\infty} (\La +\phi h))(\La^{-(2 k +1)}))_{h \la^{0}}$
that $\Ad^{-1}(P\exp
\int^{\infty}_{-\infty} (\La + \phi h))( \La^{-1})$
has diagonal elements equal to
zero, and from $\partial_{k}^{SG} (1+\phi\La^{-1})
=$ degree $-1$ part of $\Ad^{-1} (P \exp \int_{-\infty}^{x} ( \La + \phi
h))( \La^{-(2 k + 1)})$ we deduce that
$\Ad^{-1}(P \exp
\int^{x}_{-\infty} (\La + \phi h))(\La^{-1})$ is proportional to
$\La^{-1}$~; since its square equals
$\La^{-2}$, it is equal to $\La^{-1}$. These two equations are
consequences of $\partial_{k}^{SG} M_{x} =M_{x} (M^{-1}_{x}
\La^{-(2k+1)}M_{x})_{\geq 0}$ and $\partial_{k}^{SG} n_{x} =
(M_{x}^{-1} \La^{-(2k+1)} M_{x})_{< 0} n_{x}$. The first example of this
statement is $\int^{\infty}_{-\infty} (e^{2\f} -e^{-2\f}) =0$~; it is a
consequence of $\int^{\infty}_{-\infty} \partial_{x} \partial^{SG}_{0}
\f = \partial_{0}^{ SG} [\f(\infty)-\f(-\infty)]=0$ (conservation of
topological charge).]

The higher sine-Gordon flows commute with each other and
with the mKdV flows. The solution of this system is formally written
$$M_{x}(\la ; t_{n}^{SG}, t^{KdV}_{m})= \exp (\sum_{n\geq 0}
t^{SG}_{n}\La^{-(2n+1)}) g_{0} \exp (\sum_{m\geq 0} t_{m}^{KdV} \La^{2m+1}).$$
Then to extract $e^{\f}$ from this matrix, we have to take matrix
coefficients of this group element in some integrable representation. It
means that $M_{x}(\la)$ has to be replaced by a variable $\widetilde
M_{x}(\la)$ living in the central extension of the loop group, obeying
the same equations as $M_{x}(\la)$. Now the Cartan part of $\widetilde
M_{x}(\la)$ is $\exp(\f h)$ times a central term~: it will be equal to
$\langle \La_{0}| \widetilde M_{x} (t^{SG}_{n} , t_{m}^{KdV})|\La_{0} \rangle$
if $|\La_{0}\rangle$ is a level-one highest weight vector, such that
$h|\La_{0}\rangle =0$. Now, if $|\La_{1}\rangle$
is a level-one highest weight vector,
with $h| \La_{1}\rangle= | \La_{1}\rangle$, we have
$$
e^{\f(t_{n}^{SG},t_{m}^{KdV})} = {\langle \La_{1} | \widetilde M_{x}
(t_{n}^{SG} , t_{m}^{KdV})|
\La_{1}\rangle\over \langle \La_{0} | \widetilde M_{x}(t_{n}^{SG} ,
t_{m}^{KdV})|\La_{0} \rangle} .
$$
(cf. [OTU]).

In the case of the soliton solutions, we take $g_{0} = e^{Q_{1}F(z_{1})}
\cdots e^{Q_{n}F(z_{n})}$, with $[ \La^{n},F(z)] = z^{n} F(z)$. Let
$\tau_{i} (t_{n}^{SG},t_{m}^{KdV}) = \langle \La_{1} | \widetilde M_{x}
(t^{SG}_{n} ,t_{m}^{KdV}) |\La_{1} \rangle$, then $\partial^{SG}_{n} \tau_{i}
= t_{n}^{SG} \tau_{i} + \sum^{n}_{i=1} z_{i}^{-n} {\partial
\tau_{i}\over \partial Q_{i}}$, because $[\La^{-n},g_{0}]= \sum_{i=1}^{n}
z^{-n}_{i} {\partial g_{0} \over \partial Q_{i}}$, so,
$\partial^{SG}_{n} (e^{\f})=\sum^{n}_{i=1} z_{i}^{-n}{\partial\over
\partial Q_{i}} (e^{\f})$~; similarly, $\partial_{n}
(e^{\f})=\sum^{n}_{i=1} z_{i}^{n} {\partial \over \partial Q_{i}}
(e^{\f})$.
This shows~:
\noindent
\proclaim{Proposition} The mKdV and higher SG flows span a finite
dimensional family of vector fields on the multisoliton solutions.
\endproclaim

Note also that $g_{0}$ can be considered as a monodromy matrix~; from
its expression $\prod^{n}_{i=1} \exp (\sum_{k\in \ZZ} h\La^{k}
z_{i}^{-k})$ we deduce that as a function of $\la$ it has poles at
$\la=z_{i}$.

\section{2.}{The lattice sine-Gordon system.} Let as in [EF]
$x_{i},y_{i}(i\in \ZZ)$ be a system of lattice variables, with Poisson
brackets $\{ x_{i},x_{j}\}= x_{i}x_{j}, \{ y_{i},y_{j} \}= y_{i}y_{j} ,
i < j ,\{ x_{i},y_{j}\}= - x_{i} y_{j}, i\leq j$. Consider the matrix
$$
g= \prod^{0}_{i=-\infty} \pmatrix{1 & 0\cr \la x_{i} & 1\cr} \pmatrix{1
& y_{i}\cr 0 & 1\cr} . \pmatrix{1 & 0\cr 1/\la x_{0} + 1/y_{1} \cdots &
1\cr}
\pmatrix{1 & [y_{-1} +\cdots ]^{-1}\cr 0 & 1\cr}\cdot
$$
$$\cdot
\pmatrix{ \prod_{i\ge 1} {\la
x_{i}\over \la x_{i}+ 1/y_{i}\cdots}  {y_{i}\over y_{i} + 1/\la
x_{i}+\cdots}
& 0 \cr 0 &
(\prod_{i\ge 1} {\la
x_{i}\over \la x_{i}+ 1/y_{i}\cdots}  {y_{i}\over y_{i} + 1/\la
x_{i}+\cdots})^{-1}
\cr}.
$$
The lattice sine-Gordon and mKdV flows can be expressed as
$$
\partial_{n}^{SG} g = \La^{- (2 n+1)} g \qqbox{and} \partial_{n}^{KdV}
g = g h \la^{n} , n \geq 0\  ; \leqno(1)
$$
here $\La = \pmatrix{0 & 1\cr \la & 0\cr}$. Let $M$ be the matrix
$\mathop{\prod}\limits^{\infty}_{i=-\infty} \pmatrix{1 & 0\cr \la x_{i}& 1\cr}
\pmatrix{1 & y_{i}\cr 0 & 1\cr}$.

\proclaim{Proposition} The following relation is satisfied
$$
\{ M \otimes, g\} = \rho (M \otimes g)\ ,
$$
where $\rho = - 2 \sum_{i\geq 0} e_{i} \otimes f_{-i} - 2 \sum_{i\geq 1}
f_{i} \otimes e_{-i} - \sum_{i\geq 1} h_{i} \otimes h_{-i}$. [Here
$e_{i}= \pmatrix{0 & \la^{i}\cr 0 & 0\cr}, f_{i} = \pmatrix{0 & 0\cr
\la^{i} & 0\cr} , h_{i} = \pmatrix{\la^{i} & 0 \cr 0 & -\la^{i}\cr}$.]
\endproclaim

\noindent
{\bf Proof.} For the matrix elements of $M$ corresponding to the simple
roots \break $( \sum^{\infty}_{-\infty} x_{i}$ and $\sum^{\infty}_{-\infty}
y_{i})$, this statement is $\{ \sum_{-\infty}^{\infty} x_{i},
g\}=-2e_{-1}g$,  $\{
\sum_{-\infty}^{\infty} y_{i}, g\}=-2f_{0}g$, and
it is a consequence of $\{ \sum_{-\infty}^{\infty} x_{i} ,
g\}'=-2(ge_{-1}g^{-1})_{-}g$,
$\{ \sum_{-\infty}^{\infty} y_{i}, g\}'=-2(gf_{0}g^{-1})_{-}g$ which
follow from [EF] (here $\{ a,b\}' = \{ a,b\}-(\deg a)(\deg b)ab$, $\deg
x_{i}$ $= 1$, $\deg y_{i} =-1$). We then check its compatibility with $\{ M
\otimes, M\}= r^{L} (M \otimes M) - (M \otimes M)r^{R}$~: it is $(\d
\otimes 1)(\rho)= [ \rho^{23},\rho^{13}]$, (where $\d (x)= [ r, x
\otimes 1 + 1 \otimes x])$. Recall that $\rho \in \hat b_{+} \otimes \hat
b_{-}$. Let $\xi , \eta \in \hat b_{-}$, this equality means that $\langle
\rho , [\xi , \eta ] \otimes 1] = [ \langle \rho^{13} , \xi^{1} \rangle ,
\langle \rho^{23} , \eta^{2} \rangle]$~; since $[\rho , \xi \otimes
1]=\xi$ this equality is true (we have used the duality between $\hat b_{+}$
and $\hat b_{-}$). \hfill $\bull$

Let us set $H_{i} =$ constant coefficient of $\tr(\La^{-(2 i + 1)} M)$. The
family $H_{i}$ is Poisson commutative. Then
$$
\{ H_{i}, g\} = \sum_{j}- 2 d_{i-j} (f^{L}_{-j}g) - 2 a_{i-j} (e^{L}_{-j-1} g)
- ( c_{i-j} - b_{i+1-j}) (h^{L}_{-j} g)\ ,
$$
where we have set $M= \pmatrix{a(\la) & c(\la)\cr b(\la) & d(\la)\cr}$
and $a(\la) = \sum_{i\geq 0} a_{i} \la^{i}$, etc...

The Hamiltonian flows generated by the $H_{i}$ commute with the mKdV
flows (since $H_{i}$ are in the Poisson algebra generated by $b_{1}$ and
$c_{0}$, which commute to the integrals of motions)~; these flows will
be denoted $\partial_{H_{i}}$. Generally they will make sense only at
the level of formal variables (whereas $\partial_{n}^{SG} =
f^{L}_{-n} + e_{-n-1}^{L}$ will have concrete realizations), because
$a_{i-j}$ and $d_{i-j}$ will be infinite. Nevertheless, for solutions of
the system such that $x_{n}, y_{n}\mathop{\longrightarrow}\limits_{\pm
\infty} a$, we can consider as we did in the continuous case
that $a_{i}-d_{i}=b_{i}-c_{i-1}=0$
and that the linear spans of $\partial_{H_{1}}, \cdots,
\partial_{H_{n}}$ and $\partial^{SG}_{1},\cdots,
\partial^{SG}_{n}$ are the same.

Let us now show how the vector fields $\partial^{SG}_{n}$ operate on
the space of lattice variables, rapidly tending to $a$ at infinity. Let us
express those fields as
$$
\partial_{1}^{SG}(\ln x_{n})=-2\sum_{i<n}y_{i}+ 2 \sum_{i<n} x_{i}+ x_{n}=
(x_{n}-y_{n-1} )- (y_{n-1}-x_{n-1})+ (x_{n-1}-y_{n-2})\cdots,
$$
$$
\partial_{1}^{SG} (\ln y_{n}) =(y_{n} -x_{n}) - (x_{n} -y_{n-1})+\cdots,
$$
and $\partial^{SG}_{n}(\ln x_{n})$, $\partial^{SG}_{n} (\ln y_{n})$
are given by (1), where the matrices
$$
\Ad^{-1}\prod^{i}_{j=-\infty}\pmatrix{1 & 0\cr -\la x_{j} & 1\cr}
\pmatrix{1 & -y_{j}\cr 0 & 1\cr} \cdot \pmatrix{0 & \la^{-1}\cr 1 & 0\cr} ,
$$
$$
\Ad^{-1} \prod^{i}_{j=-\infty}
\pmatrix{1 & -y_{j}\cr 0 & 1 \cr}
\pmatrix{1 & 0\cr -\la x_{j+1} & 1\cr}\cdot \pmatrix{0 & \la^{-1}\cr 1 &0\cr}
$$
are understood as
$$
\lim_{N\to +\infty} \Ad^{-1} \prod^{i}_{j=-N} \pmatrix{1 & 0\cr -\la
x_{j}& 1\cr} \pmatrix{1 & -y_{j}\cr 0 & 1\cr} \cdot {1\over \sqrt{1+
{a^{2}\la\over 4}}} \pmatrix{ {a\over 2} & \la^{-1} \cr 1 & -{a\over 2}}
\leqno(3)
$$
and
$$
\lim_{N\to +\infty} \Ad^{-1} \pmatrix{1 & 0\cr -\la x_{-N} & 1\cr}
\prod^{i}_{j=-N} \pmatrix{1 & -y_{j}\cr 0 & 1\cr} \pmatrix{1 & 0\cr -
\la x_{j+1} & 1\cr} \cdot {1\over \sqrt{1+ {a^{2}\la\over 4}}}
\pmatrix{{a\over 2} & \la^{-1} \cr 1 &  - {a\over 2}} \leqno(4)
$$
(these limits are well defined, because of the assumptions on
$x_{i},y_{i}$). Then $\partial_{k+1}^{SG} (\ln x_{n}^{(\e)})$ is a
linear combination of $\sum_{2n + \e \geq 2 i_{1} +\e_{1}+k_{1} \geq
2i_{2}+\e_{2}+k_{2}\geq \cdots}
x_{i_{1}}^{(\e_{1})}\cdots x_{i_{2k}}^{(\e_{2k})}
\partial_{1}^{SG}\ln x_{i_{2k+1}}^{(\e_{2k+1})}$
where $x_{n}^{(\e)} = x_{n}$ for $\e =0$, and $y_{n}$ for $\e =1$, and
$k_{i} - k_{i-1} =0$ or $1(k_{0}=0)$.

These equations are well defined for sequences satisfying $x_{n}, y_{n}
\mathop{\longrightarrow}\limits_{n\to -\infty} a$ faster then any
${1\over n^{k}}$, and they
preserve this boundary condition. If $(x_{n},y_{n})$ satisfies the same
conditions for $n\to +\infty$, and if all coefficients of $\la^{k}$ in
$A_{N,N'} + a B_{N,N'} - D_{N,N'}$ and $B_{N,N'} -\la C_{N,N'}$,
tend to zero faster than any ${1\over N^{\ell}} {1\over N^{\prime\ell'}}$ when
$N,N' \to +\infty$, where $M_{N,N'} = \pmatrix{A_{N N'} & C_{N N'}\cr
B_{N,N'} & D_{N,N'}\cr} = \pmatrix{1 & 0\cr \la x_{-N} & 1 \cr} \cdots
\pmatrix{1 & y_{N'} \cr 0 & 1\cr}$, then the system of equations can be
rewritten with $+\infty$ replacing $-\infty$~; these conditions on the
lattice variables are
preserved by the flow. Once again, any solution to the system, such that
$x_{n}, y_{n} \to a$ and $\partial_{t_{k}} x_{n}, \partial_{t_{k}} y_{n}
\to 0$, faster than any ${1\over n^{\ell}}$ for $x\to \pm \infty$, has to
satisfy the above conditions on $A_{N,N'}$ etc... Indeed, the system has
to take the forms
$$
\partial^{SG}_{k} n = \sum_{s} a_{k,s} \la^{-s}
\pmatrix{- {a\over 2} & \la^{-1} \cr 1 &  {a\over 2}\cr}n
$$
for $n =
\pmatrix{1 & 1/(\la x_{i})\cr 0 & 1 \cr} \pmatrix{ 1 & 0 \cr
-[\la y_{i-1}\cdots]^{-1} & 1\cr}\pmatrix{t & 0\cr 0 & t^{-1}\cr}$, and
$$
\partial^{SG}_{k} n' = \sum_{s} b_{k,s} \la^{-s} \pmatrix{ {a\over 2} &
\la^{-1}\cr 1 & - {a\over 2}\cr}n'
$$
for $n' = \pmatrix{ 1 & 0\cr
1/y_{i}\cdots & 1\cr} \pmatrix{1 & -[\la x_{i-1}\cdots]^{-1}
 \cr 0 & 1\cr}^{-1}\pmatrix{t' & 0\cr 0 & t'^{-1}\cr}$,
for $|i|$ large enough. This
implies that $(M_{N,N'})^{-1} \pmatrix{- {a\over 2} & \la^{-1} \cr 1 &
{a\over 2}\cr}M_{N,N'}= \pmatrix{\cA & \cC\cr \cB & \cD\cr}$ should
satisfy $2 \cA + a \la \cC \to 0$, and $\pmatrix{1 & a\cr 0 & 1\cr}
\pmatrix{\cA & \cC\cr \cB & \cD\cr} \pmatrix{ 1 & -a\cr 0 & 1\cr} =
\pmatrix{\cA' & \cC'\cr \cB' & \cD'\cr}$ should satisfy $2 \cA' = a \cB'$,
or $2 \cA + a\cB\to 0$. So we have $M_{N,N'}^{-1} \pmatrix{ - {a\over
2} & \la^{-1} \cr 1 & {a\over 2}\cr} M_{N,N'} \to \pmatrix{- {a\over 2}
& \la^{-1} \cr 1 & {a\over 2}\cr}$ faster than any inverse polynomial~;
the conditions (2) are satisfied.

\section{3.}{Finite dimensional orbits of lattice flows}

Let $G$ be the completion of $S\ell_{2} (\CC ((\la)))$, equal to the
product $\pi^{-1}(B_{+})\times \bar\pi^{-1} (N_{-}), \pi : S\ell_{2}
(\CC [[ \la]])\to S\ell_{2}(\CC), \la \mapsto 0$ and $\bar\pi :
S\ell_{2} (\CC [[ \la^{-1}]])\to S\ell_{2} (\CC), \la^{-1} \mapsto 0$,
and let $A_{+}$ be the subgroup of $\pi^{-1}(B_{+})$ corresponding to
the Lie algebra $a_{+}=
\oplus_{k\geq 0} \CC \la^{k} ( e + \la f)$, and $H_{-}$
the subgroup of $\bar\pi^{-1}(N_{-})$, corresponding to $h_{-}=\oplus_{k \geq
0} \CC h \la^{-k}$.

There is a bijective correspondence between elements of $A_{+} \setminus
G/H_{-}$ and systems of variables $x_{i} , y_{i}, M^{\pm}_{i} , i\in
\ZZ$, $x_{i}, y_{i}$ scalars and $M^{\pm}_{i}$ elements of
Lie $\pi^{-1}(B_{+})$, conjugate to $\la^{-1} e+ f$ and such that $M_{i}^{+}
= \Ad \pmatrix{1 & y_{i} \cr 0 & 1 \cr} M^{-}_{i}$, $M^{+}_{i+1} =
\Ad\pmatrix{ 1 & 0\cr \la x_{i+1} & 1\cr} M^{-}_{i}$. To see it, we express
$A_{+}g H_{-}$ as $A_{+} n_{+} n_{-} H_{-}$ and write $n_{-} =
\pmatrix{1 & 1/\la x_{0} + \cdots\cr 0 & 1\cr}\cdot \break \cdot\pmatrix{1
& 0 \cr - [ \la
y_{-1} + \cdots ]^{-1} & 1\cr}$, and $A_{+} g w^{\pm}
(k)H_{-} = A_{+} n^{\pm}_{+} (k) n^{\pm}_{-}(k) H_{-}$, and then pose
$M^{\pm}_{k} =\Ad n_{+}^{\pm}(k)^{-1}$ $(\la^{-1}e + f)$ [where $w^{+} (k) =
(w_{0} w_{1})^{k}$, $w^{-} (k) = (w_{0}w_{1} )^{k} w_{0}]$. It is clear
what the images of the natural flows on $A \setminus G/H_{-}$ are by this
correspondence. Finite dimensional orbits of these flows are also in
correspondence~; they are also the orbits where a linear dependance
condition between some of these flows is satisfied.

Solutions $(x_{n},y_{n})$ to (1), (2), satisfying the conditions
$x_{n}(t), y_{n}(t) \mathop{\longrightarrow}\limits_{n\to \infty} a$
faster than any ${1\over n^{k}}$, for any fixed $t$, yield solutions to the
above described system on $(x_{i},y_{i}, M_{i}^{\pm})$~: it suffices to
identify $M_{i}^{+}$ and $M_{i}^{-}$ with the matrices defined in (3) and
(4).

We will now describe the finite dimensional orbits of $A_{+} \setminus
G/H_{-}$, and describe those satisfying conditions at infinity (the
soliton-like orbits).

Let $g\in G$. It will lie in a finite dimensional orbit, iff a condition
of the form $(\La^{-1}+\sum_{k\geq 0} a_{k}\La^{2k+1})g=g \sum_{k\leq N}
b_{k}h\la^{k}$ ($b_{N} \not= 0$) is satisfied. Writing $g=n_{+}b_{-}$,
this means the existence of some $x\in s\ell_{2} (\CC[\la,\la^{-1}])$,
such that $(\La^{-1}+\sum_{k\geq 0} a_{k} \La^{2k+1})n_{+} = n_{+} x$,
and $xb_{-} = b_{-} \sum_{k\leq N} b_{k} h\la^{k}$~; $x$ has to be of
the form $\La^{-1} +$ elements of homogeneous degree $\geq 0$ and $<
2N+b_{N}h\la^{N}$. The vector fields (1), (2) induce on $x$ the flows
$$
\partial^{SG}_{n} x = \left[ x , \left( {\la^{-n} \over 1+ \sum_{k\geq 0}
a_{k} \la^{k+1}} x\right)_{+} \right]
\qqbox{and}
\partial_{n}^{KdV} x =
\left[ \left( {x \la^{n} \over b_{N} \la^{N} + \sum_{k<N} b_{k} \la^{k}}
\right)_{-}
,x\right] ,\leqno(5)
$$
$n\geq 1$, keeping $a_{k}$ and $b_{k}$ fixed.

Let us indicate how to obtain $(x_{i},y_{i}, M_{i}^{\pm})$ if we
know $x(t)$. Suppose that $g(t)$ corresponds to the dot
$x_{i}$ (resp. $y_{i}$). Then $\Ad\pmatrix{1 & -y_{i-1}\cr 0 & 1\cr}
x(t)$, resp. $\Ad\pmatrix{1 & 0\cr -\la x_{i-1} & 1\cr} x(t)$ gives the
matrix $x(t)$ corresponding to the previous dot, $y_{i-1}$ and $x_{i-1}$
respectively. That these matrices again take the form allowed for
matrices $x(t)$ determines $y_{i-1}$ and $x_{i-1}$~: posing $x(t)= \pm b_{N}
h \la^{N} + a_{N-1} e \la^{N-1} + c_{N} f\la^{N} +$ lower degree terms
at dot $x_{i}$ (resp. $y_{i}$), we deduce
$$
y_{i-1}(t) = - {2b_{N} \over c_{N}}(t),\  {\rm resp.}\  x_{i-1} (t)=- {2
b_{N}\over a_{N-1}}(t)\ .
$$
The values of the other variables $x_{k}(t), y_{k}(t)$ can be defined
inductively. $M^{\pm}_{i}$ can now be obtained as follows~:
$M^{\pm}_{i}= n_{+}^{-1} \La^{-1} n_{+}$, so $M^{\pm}_{i}(t) = {1\over 1
+ \sum_{k\geq 0} a_{k} \la^{k+1}} x(t)$, if $g(t)$ corresponds to the
dot $x_{i}$ (resp. $y_{i}$).

In general, the system (5) can be solved noticing that $x(t)$ follows
an isospectral evolution, and that it corresponds to linear flows on the
Jacobian of the curve $\det (\mu - x(\la)) = \mu^{2}- \left( \sum_{k\leq
N} b_{k} \la^{k}\right)^{2} =\mu^{2} -\la^{-1}(1 + \sum_{k\geq 0} a_{k}
\la^{k+1})^{2} = 0$. If this curve is smooth, its Jacobian is
compact and the solutions $(x_{i}, y_{i},M_{i}^{\pm})$ are expected
to have a quasiperiodic behaviour. In any case, we will see that such
solutions cannot have a soliton-like behaviour; such solutions will
correspond to rational curves with double points.

\section{4.}{Soliton solutions of the lattice flow}
\subsection{4.1}{Integration of the system}

Let us determine now the $(x(t), a_{k}, b_{k})$ corresponding to a
constant solution $x_{i} = y_{i} = a$. We have for such a solution,
$N=0$ so at dot $x_{i}$, $x= \La^{-1} + b_{1} h$ and $x_{i}(t)= - 2
b_{1}$, and at dot $y_{i}, x = \La^{-1} -b_{1}h$ and $y_{i}(t)=- 2 b_{1}$~;
so $b_{1} =- {a\over 2}$~; the constant solution corresponds to the
rational curve $\mu^{2} = ({a\over 2} )^{2} + \la^{-1}$.

Let us suppose that $(x_{i},y_{i},M_{i}^{\pm})$ has soliton behaviour, and
corresponds to a finite dimensional orbit. Then by (3), (4), we have for
any fixed $t=(t_{k})$, $M_{i}^{\pm} (t) \mathop{\longrightarrow}\limits_{i\to
-\infty} (1+ {a^{2}\la\over 4})^{-1/2} \pmatrix{\pm {a\over 2} &
\la^{-1}\cr 1 & \mp {a\over 2}\cr}$ (this means that all coordinates of
$M^{\pm}_{i}(t)$ in the natural coordinate system of
$S\ell_{2}(\CC((\la)))$ has the
limit indicated). This implies that ${1\over 1+ \sum_{k\geq 0}
a_{k}\la^{k+1}}\cdot$ $\cdot x_{(x_{i}),(y_{i})}(t)$ have the same limit~;
$x_{(x_{i})}(t)$ and $x_{(y_{i})}(t)$ denote the functions $x(t)$
relative to dots $x_{i}$ and $y_{i}$. We have now $x_{(x_{i})}(t)\to {1+
\sum_{k= 0}^{\infty} a_{k}\la^{k+1}\over (1 + ({a\over 2})^{2} \la)^{1/2}}
\pmatrix{ {a\over 2} & \la^{-1}\cr 1 & - {a\over 2}\cr}$, and since
$x_{(x_{i})}(t)$ has no terms of homogeneous degree $>N$, we deduce that
${1+ \sum_{k= 0}^{\infty} a_{k} \la^{k+1} \over (1+ ({a\over 2})^{2}
\la)^{1/2}}$ is a polynomial of degree $N$, $\prod^{N}_{i=1}
(1+\g_{i}\la)$. So the spectral curve has equation
$$
\mu^{2}= (\la^{-1}+
({a\over 2})^{2})\prod^{N}_{i=1} (1 +\g_{i}\la)^{2}; \leqno(6)
$$
 it is a
rational curve. This phenomenon is also observed in [NMPZ], II, 10,1. Posing
$(\la^{-1} + ({a\over 2})^{2})\prod^{N}_{i=1} (1+ \g_{i} \la^{2}) =
\la^{-1} + \sum^{2N}_{i=0} \a_{i} \la^{i}$, we get the linear dependance
condition $\partial_{1}^{SG} + \sum_{i=1}^{2N} \a_{i}\partial_{i}^{KdV}=0$
on our solution. If we assume all $\a_{i}$'s to be real, the checking of
$x_{i}(t) \to a, y_{i} (t)\to a$ for $t_{i}\to \pm \infty$, other
$t_{i}$'s fixed will be reduced to a problem of a finite number of
flows. (Some solutions with complex $\a_{i}$'s have the same property,
e.g. 1-soliton with Arg $\g_{\e}$ irrational.) From now on we suppose
all $\a_{i}$'s real.

Let us solve (5) in this case. Let $x(\la)= \pmatrix{a(\la) & c(\la)\cr
b(\la) & - a(\la)\cr}$, and $b(\la)=\prod^{N}_{i=1} (1- {\la\over
\la_{i}})$~; then $a (\la_{i})^{2} =\mu (\la_{i})^{2}$, so
$a(\la_{i})=\mu_{i}$, $(\la_{i},\mu_{i})$ being some point of (6), lying
over $\la_{i}$. We deduce
$$
a(\la)=\e {a\over 2} \prod^{N}_{i=1} \g_{i} \prod^{N}_{i=1} (\la
-\la_{i}) + \sum^{N}_{i=1} \mu_{i} \prod^{N}_{j=1\atop j\not= i} {\la
-\la_{j}\over \la_{i}-\la_{j}} , \e = \pm 1 \ .\leqno(7)
$$
Let us consider instead of the flows (5) their linear combinations
$$
\partial_{t'_{k}} x = [ (\la^{-k}x)_{+}, x], k\geq 0.
$$
Note that all
$\partial_{t'_{n}} x$ are $0$ for $n<N$, and for $n\geq N$ they will all be
proportional to $[x_{-}, x]$. The vector fields $\partial_{t'_{k}}$ induce on
$b(\la)$ the flows ${1\over 2} \partial_{t'_{k}} b= (\la^{-k}b)_{+}
a- (\la^{-k} a)_{+}b$ (here $(\la^{i})_{+} = \la^{i}1_{i>0}$).
Evaluation at $\la_{i}$ gives ${1\over 2}\prod_{j\not= i} (1-
{\la_{i}\over \la_{j}}) {\partial_{t'_{k}} \la_{i}\over \la_{i}} =\mu_{i}
(\la^{-k} b)_{+} (\la_{i})= \mu_{i}\la_{i}\cdot$ coefficient of $\la^{k}$ in
${b(\la)\over \la -\la_{i}}= \mu_{i}
(-1)^{k+1}\mathop{\sum}\limits^{k+1}_{i_{\a}\not= i\atop
i_{\a}{\rm all\ different}}
{1\over \la_{i_{1}}\cdots \la_{i_{k}}}$, so
$$
\partial_{t'_{k}} \la_{i} = 2 {(-1)^{k+1} \mu_{i} \la_{i}\over
\prod_{j\not=i} (\la_{j} -\la_{i})} \sum_{j_{\a}\not= i\atop {\rm all
\ different}}\la_{j_{1}}\cdots
\la_{j_{N-k-1}}.
$$

Consider now the forms $\o_{s} =\la^{s} {d\la\over \mu} , s =
-1,\cdots,N-2$, and the functions $F_{s}(t') = \sum^{N}_{i=1}
\int_{P_{0}}^{(\la_{i}(\tau),\mu_{i}(\tau))} \o_{s}$ ($P_{0}$ a fixed
point on the curve). They obey
$$
\partial_{t'_{k}}F_{s}= 2(-1)^{k+1} \sum^{N}_{i=1} {\la_{i}^{s+1}\over
\prod_{j\not= i} (\la_{j}-\la_{i})} \sum_{j_{\a}\not= i\atop {\rm all
\ different}} \la_{j_{1}}\cdots \la_{j_{N-k-1}} = -2 \d_{k, s+1}\ .
$$
by a Vandermonde system computation. So $F_{s}(t')=-2 t'_{s+1} + cst$.
But
$$
F_{s} (t')= \sum^{N}_{i=1} \int^{\la_{i}}_{\la_{0}} {\la^{s}d\la\over ((
{a\over
2})^{2} + \la^{-1} )^{1/2} \prod^{N}_{i=1} (1+\g_{i} \la)} =
\sum^{N}_{i=1} \int^{\tau_{i}}_{\tau_{0}} {- 2 (\tau^{2} - ({a\over
2})^{2})^{N-s-2}\over \prod^{N}_{j=1} (\tau^{2}+\g_{j}- ({a\over
2})^{2})} d\tau\ ,
$$
with $\tau^{2} = ({a\over 2})^{2} +\la^{-1}$~; more precisely $\tau =\mu
/ \prod^{N}_{i=1} (1+\g_{i}\la)$.

Assuming all $\g_{i}$'s different, the fraction in $\tau$ is decomposed
as
$$
-  \sum^{N}_{i=1} {(-\g_{i})^{N-s-2}\over  \prod_{j\not= i}
(\g_{j}-\g_{i}) \sqrt{({a\over 2})^{2}-\g_{i}}} \left( {1\over \tau -
\sqrt{({a\over 2})^{2}-\g_{i}}} - {1\over \tau + \sqrt{({a\over
2})^{2}-\g_{i}}} \right),
$$
for $i=-1,\cdots, N-2$.
This gives
$$
\sum^{N}_{i,j=1} {(-\g_{i})^{N-s-2} \over 2 \sqrt{({a\over
2})^{2}-\g_{i}}\prod_{k\not= i}(\g_{k}-\g_{i})} \ln {\tau
_{j}-\sqrt{({a\over 2})^{2}-\g_{i}} \over \tau_{j}+ \sqrt{({a\over
2})^{2}-\g_{i}}} = t'_{s+1} + C_{s+1}\leqno(8)
$$
$C_{s+1}$ are constants.

If we pose $C(\la)= \la^{-1}\prod^{N}_{i=1} (1- {\la\over \nu_{i}})$ and
$\s_{i} = \mu(\nu_{i})/ \prod^{N}_{j=1} (1+\g_{j}\nu_{i})$, we have
similarly
$$
\sum^{N}_{i,j=1} {(-\g_{i})^{N-s-2}\over 2\sqrt{({a\over 2})^{2}-\g_{i}}
\prod_{k\not= i} (\g_{k} -\g_{i})} \ln {\tau_{j} -\sqrt{({a\over
2})^{2}-\g_{i}}\over \tau_{j}+\sqrt{({a\over 2})^{2} -\g_{i}}} =
t'_{s+1} + C'_{s+1} \leqno(8bis)
$$
\subsection{4.2.}{One-soliton solutions}

Consider the case $N=1$. $\g_{1}=\g$ is real. We find, with $\e= -1$
$$
\tau_{1} = \sqrt{({a\over 2})^{2}-\g} {1 + K e^{2t'_{0} \sqrt{({a\over
2})^{2}-\g}}\over 1 - K e^{2t'_{0} \sqrt{({a\over
2})^{2}-\g}} }
\ ,\
\s_{1} = \sqrt{({a\over 2})^{2}-\g} {1 + K'
e^{2t'_{0}\sqrt{({a\over
2})^{2}-\g}}\over 1 - K' e^{2 t'_{0}\sqrt{({a\over
2})^{2}-\g}}}\leqno(9)
$$
$K,K'$ are (complex) constants of integration, related by
$$
{K'\over K} = {{a\over 2} -\sqrt{({a\over 2})^{2}-\g}\over {a\over 2} +
\sqrt{({a\over 2})^{2}-\g}}\ .
$$
For the solution to have the properties $\la_{1} (t'_{0})\to - {1\over
\g}\ ,\ \mu_{1} (t'_{0})\to - {1\over \g}$ for $t'_{0}\to \pm \infty$,
we need $\sqrt{({a\over 2})^{2}-\g}$ to be real~; so the condition is
$\g < ({a\over 2})^{2}$. We note that the values of $\tau_{1},\s_{1}$
are independant of the sign of $\sqrt{({a\over 2})^{2}-\g}$. We compute
$$
x_{i-1}(t'_{0})= a {(1-Ke^{t''_{0}})^{2}\over (1-K' e^{t''_{0}})(1-
{K^{2}\over K'} e^{t''_{0}})} , y_{i-1}(t'_{0})= a
{(1-K'e^{t''_{0}})^{2}\over (1-Ke^{t''_{0}})(1-
{K'^{2}\over K} e^{t''_{0}})}\leqno(10)
$$
with $t''_{0} = 2 \sqrt{({a\over 2})^{2}-\g} t'_{0}$. We have
$x_{i-1}(t'_{0}), y_{i-1}(t'_{0})\to a $ for $t'_{0}\to \pm \infty$.

Replacing $\e$ by 1 in (7) amounts to replacing $a$ by $-a$. (10) can be
considered as a one-soliton solution to the system.

The values of $x_{i}(t'_{0}), y_{i} (t'_{0})$ are given by formulae
(10), with $K$ and $K'$ multiplied by $({K'\over K})^{2}$~: it means
that the lattice translation of the solution corresponds to a time
translation of
$$
\D t'_{0}= {1\over \sqrt{({a\over 2})^{2}-\g}} \ln
\left( {{a\over 2} - \sqrt{({a\over 2})^{2}-\g}\over {a\over
2}+\sqrt{({a\over 2})^{2}-\g}}\right)^{2} \leqno(11)
$$
Note that up to time shift, complex solitons depend on one variable
$(\Arg \ K)$ and real solitons on no variable.

\subsection{4.2}{Breather solutions.} Let us consider now the case
$N=2$, $\g_{2}= \bar\g_{1}$ not real. Rather than solve (7) explicitly,
let us make the following qualitative remarks. Complex solutions depend
(up to real shifts in $t'_{0}$ and $t'_{1}$) on two real variables ($\Imm
C_{0}$ up to integral multiples of $\Re {2\pi \g_{1}\over 2\sqrt{({a\over
2})^{2}-\g_{1}} (\g_{1}-\g_{2})}$ and $\Imm C_{1}$ up to integral multiples
of $\Re {2\pi\over 2\sqrt{ ({a\over 2})^{2} -\g_{1}} (\g_{1}-\g_{2})}$).
The conditions for the solution to be real are $\tau_{2}
=\bar\tau_{1}$, or $\tau_{1}$ and $\tau_{2}$ real, so
the $T_{i}=\sum^{2}_{j=1} \ln {\tau_{j} -\sqrt{({a\over
2})^{2}-\g_{i}}\over \tau_{j} + \sqrt{({a\over 2})^{2}-\g_{i}}}$ should
be conjugate for $i=1,2$ and this means that $C_{0}$ and $C_{1}$ should
be real~: again the real solutions depend on no additional variable (up
to time shifts). Let us again consider  a complex solution and let us
study its evolution for large times with $\a t'_{0} +\b t'_{1}$ fixed
($\a ,\b$ real not both zero). Since $T_{1}= 2 \sqrt{({a\over
2})^{2}-\g_{1}} (t'_{0} + \g_{2} t'_{1})$, $T_{2} = 2 \sqrt{({a\over
2})^{2}-\g_{2}} (t'_{0}+\g_{1} t'_{2})$, $\Re\ T_{1}$ and $\Re\ T_{2}$ tend to
infinity except if $(1-\g_{2} {\a\over \b})/ \sqrt{({a\over
2})^{2}-\g_{2}}$ is pure imaginary (this fixes one value of ${\a\over\b}$).
Using
$$
\tau_{1}\tau_{2}(1-e^{T_{i}})- \sqrt{({a\over 2})^{2}-\g_{i}}
(1+e^{T_{i}}) (\tau_{1}+\tau_{2})+ (( {a\over
2})^{2}-\g_{i})(1-e^{T_{i}}) = 0, i= 1,2, \leqno(12)
$$
we see that $\tau_{1}$ and $\tau_{2}$ tend to $\pm \sqrt{({a\over
2})^{2}-\g_{1,2}}$ so $x_{i}$ and $y_{i}$ tend to $a$. Let us study now
the behaviour of these solutions w.r.t. lattice periodicity. The lattice
translation corresponds to some ({\sl a priori} complex) shift in times
$(t'_{0},t'_{1}) \mapsto (t'_{0} + \D_{0} , t'_{1} + \D_{1})$. $\D_{0}$
and $\D_{1}$ should be analytic functions in $\g_{1}$ and $\g_{2}$~; in
particular, let us compute them in the case where $\g_{1}$ and $\g_{2}$
are real $(\g_{1} \not= \g_{2})$. In a region
$$
| t'_{0} +\g_{2} t'_{1} |\  \ll\ |t'_{0} +\g_{1} t'_{1}|\ ,
$$
the solution $x_{i}(t'_{0},t'_{1})$ tends to $x_{i}^{(\g_{1})}
(t'_{0}+\g_{2}t'_{1})$, where $x_{i}^{(\g_{1})}(t'_{0})$ is the soliton
solution with parameter $\g_{1}$ described before. In this situation,
the lattice translation corresponds to the shift (11), so
$\D_{0}+\g_{2}\D_{1}= {2\over \sqrt{({a\over 2})^{2}-\g_{1}}} \ln
{{a\over 2} - \sqrt{({a\over 2})^{2}-\g_{1}}\over {a\over 2} +
\sqrt{({a\over 2})^{2}-\g_{2}}}$~; we have the same equality exchanging
$\g_{1}$ and $\g_{2}$, so
$$
\D_{0} = {1\over \g_{1} -\g_{2}} \sum^{2}_{i=1} {2\g_{i}(-1)^{i}\over
\sqrt{({a\over 2})^{2}-\g_{i}}} \ln {{a\over 2} - \sqrt{({a\over
2})^{2} -\g_{i}}\over {a\over 2} + \sqrt{({a\over 2})^{2}-\g_{i}}},
$$
$$
\D_{1} = {1\over \g_{1}-\g_{2}}
\sum^{2}_{i=1} {2(-1)^{i}\over
\sqrt{({a\over 2})^{2}-\g_{i}}} \ln {{a\over 2} - \sqrt{({a\over
2})^{2}-\g_{i}}\over {a\over 2} + \sqrt{({a\over 2})^{2}-\g_{i}}}\ .
$$
We see that in the case $\g_{2}=\bar\g_{1}$, $\D_{0}$ and $\D_{1}$ are
real. Note that the times direction $(\D_{0},\D_{1})$ corresponds to a
direction where the solution tends to $a$~; we have then
$x_{i+N}(t'_{0},t'_{1})= x_{i}(t'_{0}+N\D_{0},t'_{1}+N \D_{1})\to a$
when $N\to \infty$. One can easily show that this convergence in
exponential in $N$. So the solutions found here (with
$\g_{2}=\bar\g_{1}$) have soliton-like behaviour (in the sense of 4.1) and
can be thought of as analogues of the breathers (despite the fact that
solitons are not charged).

\subsection{4.4}{$N$-soliton solutions.} Let us turn now to the general
case~: we have $p$ real poles $\g_{1},\cdots, \g_{p}, (\g_{i}< ({a\over
2})^{2})$ and $q$ pairs of complex poles $\g_{1},\bar\g_{1},\cdots,
\g_{q},\bar\g_{q}$~; all pairwise different. We will show that the
corresponding solution has soliton-like behaviour, and that it corresponds
to the scattering of $p$ solitons and $q$ breathers. Now $N= p+2q$. Let
us discuss the effect of a large translation in times along some vector
$(\a_{0},\cdots, \a_{N-1})$. (8) can be solved by
$$
\sum^{N}_{j=1}\ln {\tau_{j}- \sqrt{({a\over 2})^{2}-\g_{i}}\over
\tau_{j} + \sqrt{({a\over 2})^{2}-\g_{i}}}
= 2(-1)^{i}\sqrt{({a\over
2})^{2}-\g_{i}} \sum^{N}_{j=1}
\sum_{i_{\a_{k}}\not= i\atop {\rm all\
different}} \g_{i_{\a_{1}}} \cdots \g_{i_{\a_{j-1}}}(t'_{j-1}+ C_{j-1})
=T_{i}\leqno(13)
$$
As long as $\Re \left[ 2 \sqrt{({a\over 2})^{2}-\g_{i}} \sum^{N}_{j=1}
\sum_{i_{\a_{k}} \not= i\atop {\rm all\ different}} \g_{i_{\a_{1}}}
\cdots \g_{i_{\a_{j-1}}}
\a_{j-1}\right]$ are all non zero, all $\Re(T_{j})$ will tend to
infinity, which means that $\tau_{j}$ tend to $\sqrt{({a\over 2})^{2}
-\g_{j}}$ (up to order) and $x_{a}(t'_{i}), y_{\a}(t'_{i})$ tends to
$a$.

Let us discuss now the effect of a lattice translation. It corresponds
to some shift in times $(t'_{i}) \mapsto (t'_{i} + \D_{i}), i =
0,\cdots, N-1$. We again compute $\D_{i}$ by analytic continuation of
the case where all $\g_{i}$'s are real. Fix some index $i$. In a
region $| T_{i}| \ll |T_{k}|, k\not= i$, we have $\tau_{k} \to \pm
\sqrt{({a\over 2})^{2} -\g_{k}}$ and $x_{\a}(t'_{0},\cdots,t'_{N-1})$
tends to $x_{\a} (T_{i} + {\rm const})$. Proceeding as we did previously, we
find
$$
\D_{\a}= \sum^{N}_{i=1} {2 (-\g_{i})^{N-1-\a}\over \sqrt{({a\over
2})^{2}-\g_{i}}} \ln {{a\over 2} -\sqrt{ ({a\over
2})^{2}-\g_{i}}\over {a\over 2} + \sqrt{({a\over 2})^{2}-\g_{i}}}\ .
$$
In our case, $\D '_{\a}s$ are real. Since $\Re( 2 (-1)^{i}\sqrt{({a\over
2})^{2}-\g_{i}}\sum^{N}_{j=1} \sum_{i_{\a_{k}\not= j} \atop {\rm all\
different}}$ $\g_{i_{\a_{1}}} \cdots \g_{i_{\a_{j-1}}} $ $\D_{j})=\Re
\left( \ln {{a\over 2}-\sqrt{({a\over 2})^{2}-\g_{i}}\over {a\over
2} + \sqrt{({a\over 2})^{2}-\g_{i}}}\right)$, it is not zero, by the
assumptions. We deduce again that for any $(t'_{i}), x_{\a}(t'_{j})$ and
$y_{\a}(t'_{i})$ tend to $a$ as $\a \to \pm \infty$~; this convergence
is exponential in $\a$, as is easily shown. So the solution has
soliton behaviour.

Let us indicate the regions of large times where the solution does not
tend to $a$~: we have $|T_{i} |\ll |T_{k}|,| T_{\ell}+\bar T_{\ell}|$, for
$i \leq p$, $k\leq p$ and $k \not= 0$, $\ell \geq p+1$~; and $| T_{\ell}
+ \bar T_{\ell} | \ll | T_{k}| , | T_{\ell'} + \bar T_{\ell'}|$,
for $i\leq p , \ell, \ell' \geq p+1, \ell \not= \ell'$. They correspond
to solitons and breathers with parameters $\g_{i}, i=1,\cdots, N-2$.

Let us compute now the phase shift of the particle $\g_{i}$, due to the
scattering with other particles.

We find, when $\tau_{j}$ goes from $\sqrt{({a\over 2})^{2}-\g_{j}}$ to
$-\sqrt{({a\over 2} )^{2}-\g_{j}}$, for $j\not= i\ ,\ \D T_{i}=
\sum_{j\not= i} \ln {\sqrt{({a\over 2})^{2}-\g_{i}} - \sqrt{(
{a\over 2})^{2} -\g_{j}}\over \sqrt{({a\over 2})^{2}-\g_{i}} + \sqrt{(
{a\over 2})^{2} -\g_{j}}}$ if $i\leq p$, and $\D (T_{\ell} + \bar
T_{\ell})= \sum_{j\not= i, \bar i} 2 \Re\  \ln{{\sqrt{({a\over
2})^{2}-\g_{i}} - \sqrt{(
{a\over 2})^{2} -\g_{j}}\over \sqrt{({a\over 2})^{2}-\g_{i}} + \sqrt{(
{a\over 2})^{2} -\g_{j}}}}$ if $i>p$. So the phase shifts are the sums
of two-particle contributions~; this is a situation of elastic
scattering.

\subsection{4.5.}{Multiple-poles solutions.} Assume that the poles $\g_{0} ,
\g_{j} , \bar \g_{j} (i \leq p, j\geq p+1)$ have multiplicities
$\a_{i},\a_{j}$. Formula (8) is then replaced by
$$
\sum^{p+q}_{i=1} {1\over \prod_{k\not=i} (\g_{k} -\g_{i})^{\a_{i}
+\a_{k}}}
{1\over (\a_{i}-1)!} ({\partial\over \partial\g_{i}})^{\a_{i}-1}
\left[ {(- \g_{i})^{N-s-2} \over 2\sqrt{ ({a\over 2})^{2}-\g_{i}}}
\sum^{N}_{j=1} \ln { \tau_{j} - \sqrt{({a\over 2})^{2}-\g_{i}}\over
\tau_{j} + \sqrt{ ({a\over 2})^{2} -\g_{i}}}\right]
\leqno(14)$$
$= t'_{s+1} +cst $,
where $N= \sum^{p+q}_{i=1} \a_{i}$, because $\lim_{\g_{i}\to \g}
\sum^{k}_{i=1} {F(\g_{i})\over \prod_{j\not= i} (\g_{i}-\g_{j})} =
{F^{(k)}(\g)\over k!}$, which is solved by
$$
({\partial\over \partial\g_{i}})^{\b} \sum^{N}_{j=1} \ln {\tau_{j} -
\sqrt{ ({a\over 2})^{2} -\g_{i}} \over \tau_{j} + \sqrt{({a\over
2})^{2}-\g_{i}}} = T_{i,\b} \ ,\ \b =0,\cdots,\a_{i-1} , \leqno(15)
$$
where the $T_{i,\b}$ are linear combinations of the times $t'_{s+1}$. The
effect of a lattice translation and the large time behaviour of this
system are studied in the same way as for the simple poles solutions~:
so multiple poles again define solitonic solutions, whose scattering can
be computed similarly.

We note two main differences with the continuous case~: absence of
topological charge, and no fermionic nature of the solitons.

\subsection{5. Periodic solutions of the lattice flow.}{}

We now examine the possibility of finite dimensional orbits of the
lattice flow yielding periodic (in $i$) solutions $x_{i}(t),y_{i}(t)$. Let
$\nu$ be this period. The condition for periodicity is the existence of
an element $b$ in $\widehat N_{+}$, of the form $\pmatrix{*
\la^{\nu-1}+\cdots *\la^{\nu-1}+\cdots \cr *\la^{\nu}+ \cdots
*'\la^{\nu}+\cdots  \cr}$, $*' \not= 0$ (for example), commuting to
$x_{(x_{i})}(t)$.
This condition is preserved by the flows~: $b$ has to follow the
equations $\partial_{t_{n}}^{SG} b = [ b,({\la^{-n}\over 1+ \sum
a_{k}\la^{k+1}}x )_{+}] = [ ( {\la^{-n} \over 1+ \sum
a_{k}\la^{k+1}}x)_{-},b]$ and $\partial_{t_{n}}b = [ ({ x\la^{n}\over
b_{N}\la^{N}+\sum_{k< N}b_{k}\la^{k}})_{-} ,b]=- [({ x\la^{n}\over
b_{N}\la^{N}+\sum_{k< N}b_{k}\la^{k}})_{+},b]$ which clearly
preserve its form.
Writing $b= P+Q x$, $P, Q$ rational fractions in
$\la$, the conditions on $P$ and $Q$ are~: $P$ polynomial, $Qx$
polynomial and $P^{2}+ Q^{2}\det x =1$.
Thus the condition is the existence of a meromorphic function on the
spectral curve, with only zeroes and poles at the points at infinity
$\infty^{\pm}$. The order of this zero (or pole) is $\nu$, and we see
that we have $\nu(\infty^{+}-\infty^{-})=0$ is the Jacobian of the curve;
$\nu$ is also the lattice period. Thus the periodicity condition is that
the difference of the infinity points is torsion is the Jacobian of the
spectral curve.

We compute now the dimension of the space of
periodic solutions among all spectral curves of degree $N$. Writing $\det x = -
\la^{-1}\prod^{2N+1}_{i=1} (1- {\la\over \la_{i}})$ and $Q= {R\over S},
R,S$ polynomials of minimal degree, we find that $\la$ should divide
$R$. We find also that $S$ should divide $\prod^{2N+1}_{i=1} (1 -
{\la\over \la_{i}})$, and also $x$~; but we can restrict ourselves without
loss of generality, to the case where no factor of $\prod^{2N+1}_{i=1}
(1- {\la\over \la_{i}})$ divides $x$. Let now $R=\la R_{0}, S=1$~; the
condition $Qx$ polynomial is now satisfied, and we need only to satisfy
$P^{2} = 1 + R_{0}^{2} \la \prod^{2N+1}_{i=1} (1 - {\la \over
\la_{i}})$. The degrees of $P$ and $R_{0}$ are $\nu$ and
$\nu-(N+1)$~; let us write ${P\over R_{0}} = \left[ \la
\prod^{2N+1}_{i=1} (1 - {\la \over \la_{i}})\right]^{1/2} \left[ 1 +
{K\over \la^{2\nu-2(N+1)}} + O (\la^{-2\nu-2(N+1)})\right]$~; we obtain
${P_{0}\over R_{0}} = ( [ \la \prod^{2N+1}_{i=1} (1 - {\la \over
\la_{i}})]^{1/2})_{-} + {k'\over \la^{2\nu - (N+1)}}$, for $P_{0}$ the
remainder of the division of $P$ by $R_{0}$ and $K' = K (- \prod_{i=1}^{2N+1}
\la_{i})^{-1/2}$. The space of possible fractions has dimension $2 (\nu
-N-1)$, and $2 \nu - (N+2)$ equations must be satisfied in it~; this
imposes $N$ conditions on the $(\la_{i})_{i=1,\cdots,2N+1}$. It would be
interesting to see if some natural Poisson brackets on the $(\la_{i})$
can be introduced, and what kind of submanifold the periodic solutions
will be in the space of all solutions (or of hyperelliptic curves) with
respect to this Poisson geometry.

\vskip 1truecm
\noindent
{\bf References}
\bigskip
\item{[DS]} V.G. Drinfeld, V.V. Sokolov, {\sl Lie algebras and equations
of Korteveg-de Vries type,} Jour. Sov. Math., 30
(1985), 1975-2036.
\medskip
\item{[EF]} B. Enriquez, B.L. Feigin,
{\sl Integrals of motions of classical lattice sine-Gordon system,} to
appear in Teor. Mat. Fiz.
\medskip
\item{[OTU]} D.I. Olive, N. Turok, J.W.R. Underwood, {\sl
Affine Toda solitons and vertex operators,} Nucl. Phys. B409, 409
(1993).
\medskip
\item{[NMPZ]} S.P. Novikov, S.V. Manakov, L.P. Pitaevskii, V.E. Zakharov,
{\sl Theory of solitons,} Contemp. Sov. Math., Consultants Bureau, New
York (1984).

\adresse{Centre de math\'{e}matiques\cr
URA 169 du CNRS\cr
Ecole Polytechnique\cr
91128 Palaiseau Cedex\cr
France\cr}

\bye